# Partial Cross-Correlation of D-Sequences based CDMA System

Sandeep Chalasani

**Abstract**

Like other pseudorandom sequences, decimal sequences may be used in designing a Code Division Multiple Access (CDMA) system. They appear to be ideally suited for this since the cross-correlation of d-sequences taken over the LCM of their periods is zero. But a practical system will not, in most likelihood, satisfy the condition that the number of chips per bit is equal to the LCM for all sequences that are assigned to different users. It is essential, therefore, to determine the partial cross-correlation properties of d-sequences. This paper has performed experiments on d-sequences and found that the partial cross-correlation is less than for PN sequences, indicating that d-sequences can be effective for use in CDMA.

**Introduction**

Code Division Multiple Access (CDMA) system is widely used in mobile communications and other advanced services [1]. Currently, PN sequences are used to generate the CDMA signals. The peak cross correlation values of the PN sequences is over 0.17, with negative peaks nearing 0.34, indicating that they can generate considerable cross-talk [1]. Even the use of Gold sequences [1], which are generated using PN sequences, helps matters only to a limited extent for their peak cross-correlation remains over 0.12.

The use of decimal sequences for random number generation is extensively discussed in several papers [2-12]. Binary decimal sequences can be generated by using the formula [4]:

$$a(i) = 2^i \bmod p \bmod 2$$

where p is a prime number. Cross correlation is a standard method of estimating the degree to which two series are correlated. Consider two sequences u(i) and c(i), the cross-correlation can be given by

$$G(k) = \frac{1}{P} \sum_{i=1}^{P} u(i) c(i+k)$$

where P is the LCM of the periods of the sequences and k can be between 0 to P.

We know that the cross correlation value of two different d-sequences where the period is equal to the LCM is zero [2].

In a d-sequence based CDMA system, it is not necessarily a given that the chip count per bit for each user is equal to the LCM. This leads to two interesting questions:

1. What is the partial cross-correlation between two d-sequences?
2. What are some examples of sets of d-sequences with a convenient LCM value?

In this brief note, we consider both these questions.

## Partial Cross-Correlation

To define the context for speaking about the partial cross-correlation function, we consider Figure 1 which represents the basic operation of a spread-spectrum system in which the binary data stream, b(t), is multiplied by the spreading random sequence, c(t), to give us the output spread-spectrum sequence, m(t):

$$m(t) = c(t)b(t)$$

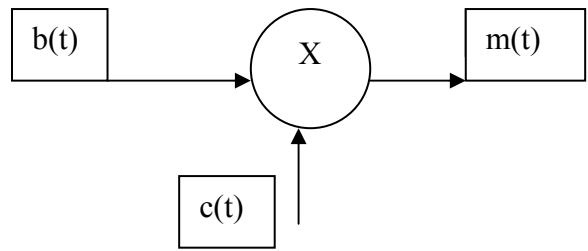

Figure 1. The basic spread-spectrum system, where c(t) is the random sequence

In computing the partial cross-correlation, we consider a value of N over which the correlation is performed that is less than the LCM of the periods of the two sequences:

$$G(k) = \frac{1}{N}\sum_{i=1}^{N} u(i)c(i+k)$$

where $1 \leq N \leq P$

**Example 1:** Consider two prime numbers $p_1 = 11$ and $p_2 = 19$

The decimal sequence for $p_1$ is 000101110, and the decimal sequence for $p_2$ is 000011010111100010

The LCM, for the periods of 10 and 18, is 90. The partial cross correlation values up to N=(P-10) and for K= 25 is given in the table below.

| | | | |
|---|---|---|---|
| 0 | 0.012500 | 13 | 0.00000 |
| 1 | 0.062500 | 14 | 0.10000 |
| 0. | 0.025000 | 15 | 0.02500 |
| 3 | -0.012500 | 16 | -0.062500 |
| 4 | 0.012500 | 17 | 0.07500 |
| 5 | -0.050000 | 18 | 0.112500 |
| 6 | 0.087500 | 19 | 0.012500 |
| 7 | 0.050000 | 20 | 0.062500 |
| 8 | -0.075000 | 21 | 0.025000 |
| 9 | 0.012500 | 22 | -0.012500 |
| 10 | 0.025000 | 23 | 0.012500 |
| 11 | -0.062500 | 24 | 0.050000 |
| 12 | 0.062500 | 25 | 0.087500 |

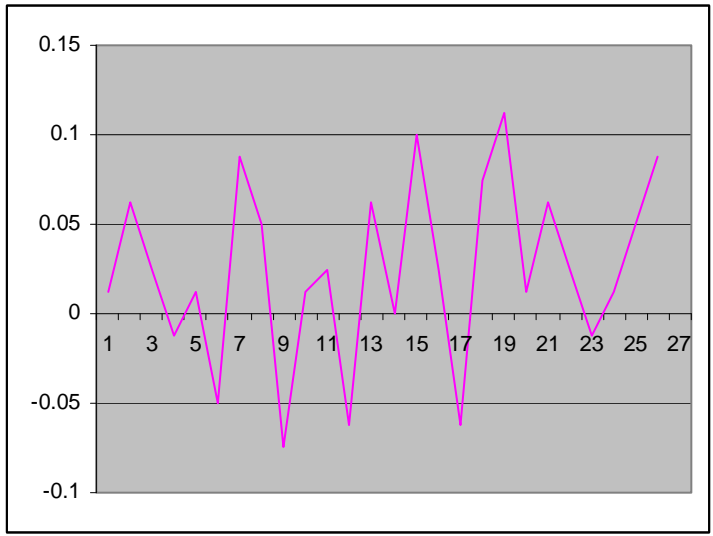

Fig.2 Partial Cross-correlation for Example 1

The maximum absolute value of the partial cross-correlation function is 0.11. This function lies most often within the range -0.5 to 0.5.

**Example 2.** Consider two prime numbers $p_1= 41$ and $p_2=17$

The decimal sequences will be 000001100011111001110000011000 and 000011110000111

LCM will be 180 and the partial cross correlation values up to N= (P-70) and for K= 32 is given in the table below.

| | | | |
|---|---|---|---|
| 0 | 0 | 16 | 0.1 |
| 1 | -0.06667 | 17 | 0 |
| 2 | 0 | 18 | -0.06667 |
| 3 | -0.03333 | 19 | 0 |
| 4 | 0.1 | 20 | -0.03333 |
| 5 | 0.033333 | 21 | 0.1 |
| 6 | -0.03333 | 22 | 0.033333 |
| 7 | 0.033333 | 23 | -0.03333 |
| 8 | 0.033333 | 24 | 0.033333 |
| 9 | -0.1 | 25 | 0.033333 |
| 10 | 0.033333 | 26 | -0.1 |
| 11 | 0.1 | 27 | 0.033333 |
| 12 | 0.166667 | 28 | 0.1 |
| 13 | 0.1 | 29 | 0.166667 |
| 14 | -0.03333 | 30 | 0.1 |
| 15 | 0.033333 | 31 | -0.03333 |

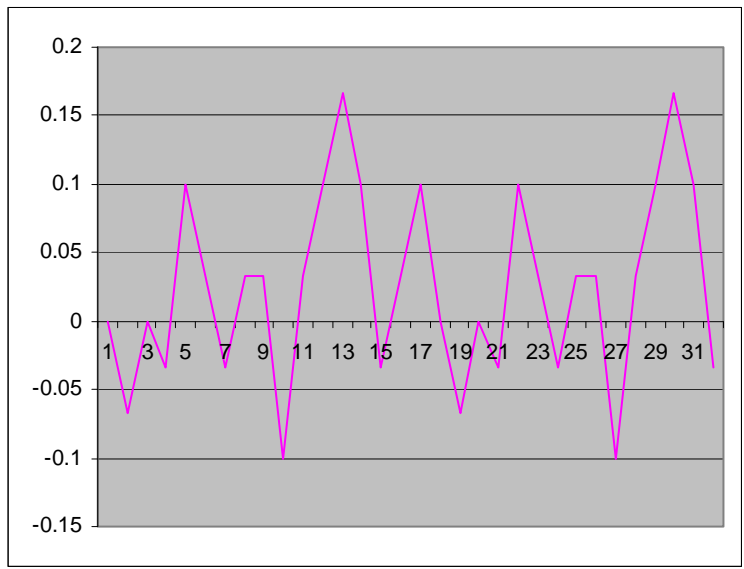

Fig.3 Partial Cross-correlation for Example 2

Although the partial cross-correlation function is over a larger range, it is still quite small. The absolute value is generally less than 0.1.

**Example 3.** For prime numbers P1=17 and p2=129 .The partial cross-correlation graph turns out to be as shown in Figure 4.

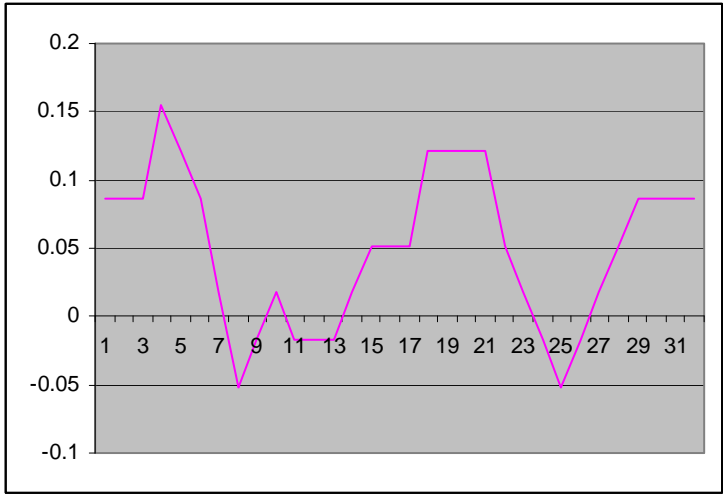

Fig.4 Partial Cross-correlation for d-sequences for 17 and 129

As in the previous examples, the cross-correlation value is mostly less than 0.1.

**Good Set of Primes**

Having a good set of primes so that the LCM of the P-1 is minimized is always desirable, a few set of primes whose LCM is minimized are

| Prime Number Sets | LCM of P-1 |
| --- | --- |
| 661,199, 397, 463, 331,101, 61 | 69300 |
| 541, 487, 379, 433, 271, 109, 163 | 136800 |
| 701, 101, 11, 71, 211, 491, 631, 281 | 88200 |
| 17, 31, 37, 61, 71, 541, 379, 433,271, 109,163 | 45360 |
| 811, 487, 163 | 2430 |
| 881, 89, 41, 17 | 880 |
| 331, 199, 11, 31, 67, 991 | 990 |
| 11, 31, 251, 31, 151, 751 | 750 |

If one were to use one of these sets then, of course, there would be guarantee of zero cross talk.

## Conclusions

We conclude that the partial cross-correlation of d-sequences is quite small and less than the corresponding values for PN sequences. This makes them competitive to PN sequences for use in the design of a CDMA system.

As a designer one has the choice of using chips per bit that satisfy the LCM constraint. We have given some random sets of primes that could be used. But the problem of designing good sets of d-sequences remains to be addressed. One would also need to relate the size of LCM to the cost and complexity of the CDMA system.